\documentclass[a4paper,twoside]{article}

\usepackage{epsfig}
\usepackage{subfigure}
\usepackage{calc}
\usepackage{amssymb}
\usepackage{amstext}
\usepackage{amsmath}
\usepackage{amsthm}
\usepackage{multicol}
\usepackage{pslatex}
\usepackage{apalike}

\usepackage{eurosym}
\PassOptionsToPackage{hyphens}{url}\usepackage{hyperref}
\usepackage{xcolor}

\usepackage{SCITEPRESS}     % Please add other packages that you may need BEFORE the SCITEPRESS.sty package.

\begin{document}
\title{Preventing Hospital Acquired Infections Through a Workflow-Based Cyber-Physical System}

\author{\authorname{Maria Iuliana Bocicor\sup{1,2}, Arthur-Jozsef Molnar\sup{1,2} and Cristian Taslitchi\sup{3}}
\affiliation{\sup{1}SC Info World SRL, Bucharest, Romania}
\affiliation{\sup{2}Faculty of Mathematics and Computer Science, Babes-Bolyai University, Cluj-Napoca, Romania}
\affiliation{\sup{3}Faculty of Automatic Control and Computers, University Politehnica of Bucharest, Bucharest, Romania}
\email{\{iuliana.bocicor, arthur.molnar\}@infoworld.ro, cristian.taslichi@gmail.com}
}


\abstract{Hospital acquired infections (HAI) are infections acquired within the hospital from healthcare workers, patients or from the environment, but which have no connection to the initial reason for the patient's hospital admission. HAI are a serious world-wide problem, leading to an increase in mortality rates,  duration of hospitalisation as well as significant economic burden on hospitals. Although clear preventive guidelines exist, studies show that compliance to them is frequently poor. This paper details the software perspective for an innovative, business process software based cyber-physical system that will be implemented as part of a European Union-funded research project. The system is composed of a network of sensors mounted in different sites around the hospital, a series of wearables used by the healthcare workers and a server side workflow engine. For better understanding, we describe the system through the lens of a single, simple  clinical workflow that is responsible for a significant portion of all hospital infections. The goal is that when completed, the system will be configurable in the sense of facilitating the creation and automated monitoring of those clinical workflows that when combined, account for over 90\% of hospital infections.}

\onecolumn \maketitle \normalsize \vfill

\section{\uppercase{Introduction}}
\label{sec:introduction}

Hospital acquired infections (HAI) or nosocomial infections are defined as infections \emph{''acquired in hospital by a patient who was admitted for a reason other than that infection. An infection occurring in a patient in a hospital or other healthcare facility in whom the infection was not present or incubating at the time of admission. This includes infections acquired in the hospital but appearing after discharge, and also occupational infections among staff of the facility''} \cite{WHO02}. Thus, in addition to decreasing the quality of life and increasing mortality rates, the duration of hospitalisation as well as the costs of medical visits for patients, HAI represent a direct occupational risk for healthcare workers, which significantly affects costs and has the potential of creating personnel deficits in case of an outbreak.

Existing research shows that HAI are prevalent across the globe, regardless of geographical, political, social or economic factors \cite{WHO02}, \cite{WHO10}. Even more compelling is the fact that while the sophistication of medical care is constantly increasing, reported HAI rates have not seen meaningful decrease \cite{tikhomirov87}, \cite{coello93}, \cite{WHO10}, \cite{ecdc15}. 

Studies undertaken between 1995 and 2008 in several developed countries have revealed infection rates between 5.1\% to 11.6\% \cite{WHO10}; data from lesser developed countries is in many cases limited and deemed of low quality. According to the European Centre for Disease Prevention and Control, approximately 4.2 million HAI occurred in 2013 alone in European long-term care facilities, the crude prevalence of residents with at least one HAI being 3.4\%; this translates to more than 100 thousand patients on any given day \cite{ecdc15}. The total cases of HAI in Europe amount to 16 million extra  hospitalisation days, 37.000 attributable deaths and an economic burden of \euro 7 billion in direct costs \cite{WHO10}. A study conducted in the United Kingdom \cite{plowman00} concluded that patients who developed HAI stayed in hospital 2.5 times longer and the hospital costs (for nursing care, hospital overheads, capital charges, and management) tripled. In the USA, in 2002 alone, 1.7 million patients were affected by HAI and the annual economic impact was approximately US\$6.5 billion in 2004. Besides all these resource expenses, what is even worse is that HAI are responsible for an increase in mortality rates, as these infections lead to death in 2.7\% of cases \cite{WHO09}. A more recent study \cite{magill14} conducted in 183 US hospitals revealed that 4\% of the patients had one or more HAI, which lead to an estimated number of 648.000 patients with HAI in acute care hospitals in the US, in 2011. According to a recent study \cite{ling15}, the prevalence of HAI in Southeast Asia between 2000 and 2012 was 9\%, the excess length of stay in hospitals of infected patients varied between 5 and 21 days and the attributed mortality was estimated between 7\% and 46\%. In Canada, more than 220.000 HAI result in 8000 deaths a year, making infections the fourth leading cause of death in the country, with \$129 million in extra  costs incurred in 2010 \cite{canada13}, \cite{canada14}. In Australia, there is an estimated number of 200.000 cases of HAI per year, resulting in 2 million hospitalization days \cite{cruickshank08}.

The most common target sites of HAI are the urinary and respiratory tracts and areas involved in invasive procedures or catheter insertion areas. While the methodology for prevention exists, it is often ignored due to lack of time, unavailability of appropriate equipment or because of inadequate staff training. Research shows that the most important transmission route for HAI are members of staff coming into contact with patients or contaminated equipment without following proper hygiene procedures \cite{hammer}.

During the twentieth century, several specific measures were taken to prevent the occurrence and spread of infections, which have been translated into a series of instructions for controlling the vectors that propagate infection as well as to properly manage outbreaks and epidemics. Originally, these instructions were established as guidelines for healthcare workers, later they were also transformed into prevention rules included in the documentation concerning workplace safety, and in recent years they were incorporated into various software or cyber-physical solutions to monitor and ensure compliance. An unquestionable benefit and substantial improvement in this respect has been brought by the Internet of Things (IoT) technologies, which are currently employed in healthcare and many other areas of life. Since the use of various IoT monitoring systems, significant improvement has been observed regarding compliance to hygiene regulations and prevention standards, as well as decreases in the rates of infections. 

In this paper we present an IoT-based cyber-physical system that targets  HAI prevention, on which development has started under funding from the European Union. Integrating a network of sensors that monitor clinical workflows and ambient conditions with monitoring software, the system will provide real-time information and alerts. The system will monitor general processes known to affect HAI spread such as cleaning and equipment maintenance, together with clinical processes at risk for HAI such as catheter insertion, postoperative care or mechanical ventilation. In this paper, we will illustrate the system using a clinical workflow that is often involved in HAI transmission, together with a motivating example that details a software perspective regarding how the system ensures compliance to established preventive guidelines.

\section{\uppercase{Related Work}}
\label{sec:related_work}

Several automated solutions have been implemented to avert and reduce HAI and their impact. The underlying idea used by several of these systems is continuous monitoring of healthcare workers' hand hygiene and real-time alert generation in case of non-compliance with established guidelines. The final purpose is to modify human behaviour towards better hand hygiene compliance. This is usually accomplished using wearable devices worn by healthcare workers, which interact with sensors placed in key hospital locations in order to record personnel activities and hand hygiene events. When a hand hygiene event is omitted or performed, the device provides visual, auditory or haptic notification. The IntelligentM \cite{intelligentM} and Hyginex \cite{hyginex} systems use bracelet-like devices that are equipped with motion sensors to ensure that hand sanitation is correctly performed. Biovigil technology \cite{biovigil} and MedSense \cite{medsense} are designed for the same purpose, only in these cases bracelets are replaced with  badges worn by the healthcare workers. The Biovigil device uses chemical sensors to detect whether hand hygiene is in accordance with established standards. The systems can be configured to remind clinicians to disinfect their hands before entering patient rooms, or before procedures such as intravenous infusions or catheter insertion. Furthermore, these systems record hygiene events, centralise them and enable analysis, visualisation and report generation. Unlike the solutions presented so far, the SwipeSense \cite{swipe_sense} system employs small devices that are alcohol-based and easy to use gel dispensers which can be worn by medical personnel. This  exempts healthcare workers from interrupting their activities in order to walk to a sink or a disinfectant dispenser \cite{simonette13}. 

A different type of system that is of assistance in the fight against infection is Protocol Watch \cite{protocol_watch}, a decision support system used to improve compliance with the \emph{''Surviving Sepsis Campaign''} international guidelines \cite{ssc} for the prevention and management of sepsis. Protocol Watch regularly checks certain medical parameters of the patients \cite{protocol_watch}, its main goal being to reduce the time period between the debut of sepsis (the moment when it is first detected) and the beginning of treatment. If the system detects that certain conditions that may cause sepsis are met, it alerts the medical staff and indicates which tests, observations and interventions must be performed, according to prevention and treatment protocols.

A different approach for the prevention of infection is taken by Xenex \cite{xenex}: the Xenex \emph{''Germ-Zapping Robot''} can disinfect a room by using pulses of high-intensity, high-energy ultraviolet light. The robot must be taken inside the room to be disinfected and in most cases, the deactivation of pathogens takes place in five minutes. After disinfection, the room will remain at a low microbial load, until it is recontaminated by a patient, healthcare worker or through the ventilation system.

Another relevant issue regards the problem of identifying control policies and optimal treatment in infection outbreaks, as introduced in \cite{curtis13}. The authors propose a comprehensive approach using electronic health records to build healthcare worker contact networks with the objective of putting into place efficient vaccination policies in case of outbreaks. Relevant software systems  developed to also improve treatment policies in case of outbreaks and epidemics are \emph{RL6: Infection} \cite{rl6} - a software solution developed to assist hospitals in the processes of controlling and monitoring infections and outbreaks, and Accreditrack \cite{accreditrack} - a software system designed to ensure compliance with hand hygiene guidelines and to verify nosocomial infection management processes. 

\section{\uppercase{Motivating Example - Hand Hygiene}}
\label{sec:example}

\begin{figure}[b]
\centering
\includegraphics[width=1\linewidth]{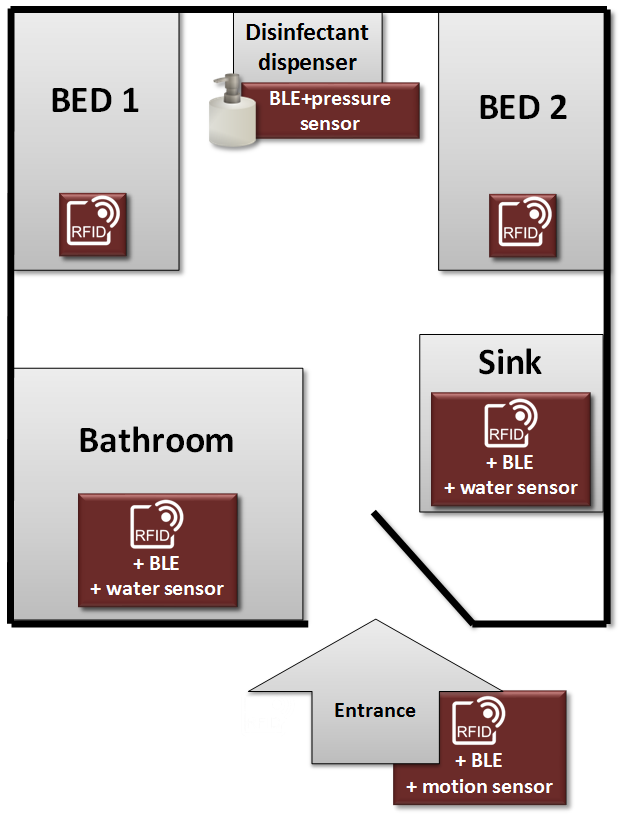}
\caption{A typical hospital room. Figure adapted from \cite{shhedi15}}
\label{fig:hospital_room}
\end{figure}

Research shows that most cases of hospital infection are tightly connected to certain clinical workflows \cite{WHO02}. Thus, we propose a cyber-physical system which monitors workflows considered relevant in HAI propagation. Although there are other ICT automated solutions that target the prevention of HAI, there are no other systems that use the clinical workflow monitoring based approach, to our knowledge. In order to successfully prevent infection and outbreaks, the system must focus on the typical sites where these infections usually occur (urinary and respiratory tracts, sites of surgery or invasive procedures \cite{WHO02}), as well as track workflows that are not infection-site specific, but which have a significant contribution to HAI rates, such as hand hygiene and transmission from the environment.

A hardware-centric motivating example of the proposed system was presented in \cite{shhedi15}. The authors of  \cite{shhedi15} focus on the hardware components of the system, namely types and location of sensors used to monitor clinical activities and the wearable technology employed by the medical staff for identification, monitoring and alerting. The purpose of our paper is to illustrate the software-side of the proposed cyber-physical system using a motivating example based on one of the most relevant, and often occurring clinical workflows. 

\begin{figure*}[t]
\centering
\includegraphics[width=1\textwidth]{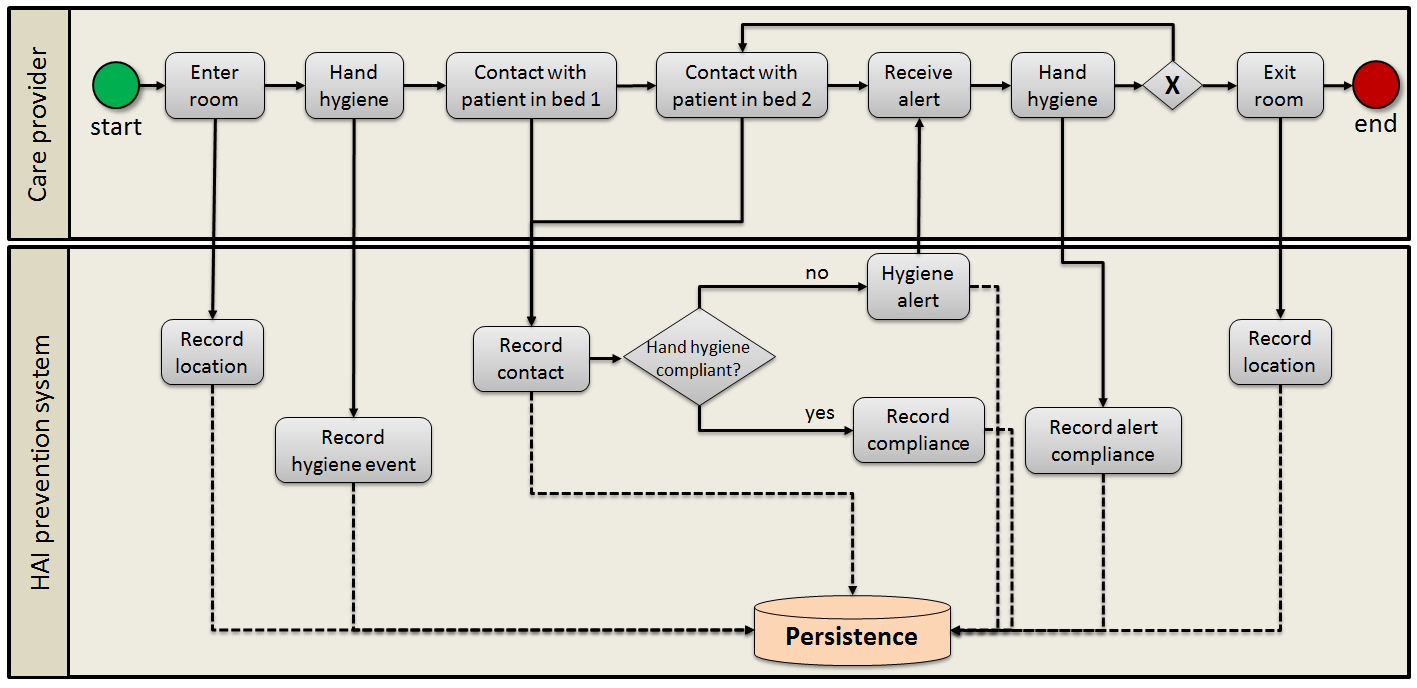}
\caption{Example of hand-hygiene relevant clinical workflow (simplified for readability).}
\label{fig:hand_hygiene}
\end{figure*}

On the software side, the system will model and encode clinical workflows using the Business Process Modelling and Notation (BPMN) \cite{bpmn} standard and it will be compatible with leading medical informatics standards such as HL7 V3 \cite{hl7_v3}, thus allowing seamless interconnection to hospital infrastructure via HL7-compliant Hospital Information Systems (HIS). In addition, it will integrate a network of hardware sensors, able to identify ongoing clinical processes, provide location information and track the use of hospital equipment and materials. Healthcare workers will use wearable devices that continuously monitor their location and activity. The central hub of the system will be a server side engine able to load and execute BPMN-based workflows. Integration with the HIS enables the retrieval of key information regarding patients, such as admissions, transfers and discharges as well as records of past of planned invasive interventions, which can be used to determine the level of risk. Hence, infections and outbreak management could be improved so that in case of suspicion, the locations of previous admissions, as well patients of members of staff considered at risk can be contacted.

The modelled clinical workflows will be executed using a BPMN engine. When the workflow leads to a state that is a risk of HAI, the engine will generate alerts that are received directly by the involved healthcare workers. In this paper we illustrate the interplay between the hardware and software components of the proposed system using a motivating example based on hand hygiene, which remains one of the most common pathways of HAI transmission. We must note that the provided example is only used to portray the workflow-based system, and that the final implementation will allow the creation and monitoring of a variety of clinical workflows, that will be executed by the BPMN engine whenever necessary.

We illustrate our motivating example using a typical scenario: a healthcare worker enters a room with two beds, and interacts with both inpatients before exiting. Both common as well as highly relevant to hand hygiene, our example is illustrated in Figure \ref{fig:hand_hygiene}, using simplified BPMN-like notation to reduce the number of decision points and emphasize readability.

According to established hand hygiene guidelines \cite{five_moments}, upon entering a patient room, workers must perform hand disinfection. If this procedure is skipped or performed inadequately (e.g. without disinfectant, shorter washing time than recommended), the system will generate an alert to warn the clinician about the detected non-compliance. After each patient contact, and before leaving the room, hand disinfection should again be correctly performed. All these events and alert reports will be persisted to enable later analyses, such as identification of an outbreak's patient zero and route of transmission.

Figure \ref{fig:hospital_room}, which was adapted from \cite{shhedi15} illustrates a typical patient room, with two beds, a sink and a bathroom. As soon as the healthcare worker enters it, the Radio-Frequency Identification (RFID) tag and motion sensor combination detect this and identify them. The system records and interprets the data received from the sensors and the BPMN engine starts a new instance of the relevant workflows, including the one for hand-hygiene. According to the hand hygiene workflow, the clinician should perform the hand disinfection procedure before going near a patient and before leaving their surrounding. In our example, this is achieved using the sink or the bathroom sink and the disinfectant dispenser. All of them have inexpensive sensors together with RFID tags and Bluetooth Low Energy transceivers; their role is to provide input to the workflow engine. The workflow engine records received data into the persistent repository. Furthermore, by running the workflow, the software engine ascertains that hand hygiene guidelines were observed. In our example, the healthcare worker performs hand disinfection before contact with the first patient, which is recorded by the sensors integrated with the sink and the disinfectant dispenser. However, the worker can move to the second patient directly, as shown in Figure \ref{fig:hand_hygiene}. In this case, the system records their proximity to the second bed via RFID; if a disinfection event that is compliant with guidelines was not recorded before the contact, the system interprets this as non-compliance, and emits an alert that is recorded and received by the healthcare worker through their wearable device. Once they become compliant by undergoing hand disinfection, they can resume contact with the patient. Information that is persisted is planned to be reused at later dates and in the context of more than one workflow, including at least all the workflows active during that time. As an example, the room entry and exit events from Figure \ref{fig:hand_hygiene} can be used for finding the source of an outbreak, or tracking its propagation. In our example, the current instance of the hand hygiene workflow ends once the healthcare worker exists the patient room.

\section{\uppercase{Conclusions}}
\label{sec:conclusions}

This paper is centred on technology-driven preventive measures for a serious public health issue, namely hospital acquired infections. While most transmission routes are well understood and standard guidelines exist to curb them, available research shows that in practice, they are either not applied, or not applied thoroughly. As such, we propose an innovative technological solution for preventing HAI and outbreaks. Our proposed approach will employ a standards-compliant, workflow based, cyber-physical system able to monitor and enforce compliance with clinical workflows that are associated with over 90\% of HAI instances, in accordance with current guidelines and best practices.

We have illustrated an overall picture of the system using a software workflow perspective through a motivating example based on hand-hygiene guidelines. However, our aim is to enable the creation, configuration and execution of location-specific clinical workflows using BPMN-like notation. As existing research shows that infections within the urinary tract, respiratory tract, surgical sites, or skin and soft tissue \cite{WHO02} account for over 75\% of the total number, we aim to also evaluate the system using workflows related to those sites. Another category is represented by workflows that are not infection-site specific, but which represent transmission vectors for numerous types of infections: hand hygiene (such as presented in this paper) and transmission from the environment.

Another future functionality, that currently sits beyond the scope of our research is a graphically interactive, configurable component that allows the management of the monitored workflows, by taking into account the specific infrastructure of the deployed sensors and of the unit of care.  Furthermore, additional components that can be added to the system include  advanced analysis tools for the gathered data that allow building risk maps and contact networks enabling new ways of pinpointing elusive reasons of hospital infection that occur even when best practices are adhered to.

In addition to the software-based functionalities, another issue that must be addressed is related to the legal aspects regarding the deployment of such a system, more precisely those regarding data protection. Considering that the system involves tracking of patients and healthcare workers alike, the software side of such systems must remain compliant with regulations regarding the protection of highly-sensitive personal data.
% ------------------------------------------------------------------------------------------------------------------------------------------------

\section*{\uppercase{Acknowledgement}}
This work was supported by a grant of the Romanian National Authority for Scientific Research and Innovation, CCCDI UEFISCDI, project number 9831\footnote{https://www.eurostars-eureka.eu/project/id/9831}.

%\vfill
\bibliographystyle{apalike}
{\small
\bibliography{example}}

\vfill
\end{document}